\documentclass{svjour2}                    
\smartqed  
\usepackage{graphicx}
\journalname{jltp}
\begin{document}

\title{Electronic coolers based on superconducting tunnel junctions: fundamentals and applications}
\subtitle{}

\author{H. Courtois \and F. W. J. Hekking \and H. Q. Nguyen \and C. B. Winkelmann}
\institute{H. Courtois \and C. B. Winkelmann \at
Universit\'e Grenoble Alpes, Institut NŽ\'eel, Grenoble, France \\ CNRS, Institut NŽ\'eel, Grenoble, France \\ \email{herve.courtois@neel.cnrs.fr,
\and H. Q. Nguyen \at Universit\'e Grenoble Alpes, Institut NŽ\'eel, Grenoble, France \\ CNRS, Institut NŽ\'eel, Grenoble, France \\ O.V. Lounasmaa Laboratory, Aalto University, Helsinki, Finland
\and F. W. J. Hekking \at
Universit\'e Grenoble Alpes, LPMMC, Grenoble, France \\ CNRS, Institut NŽ\'eel, Grenoble, France}}
\date{Received: date / Accepted: date}
\maketitle

\begin{abstract}
Thermo-electric transport at the nano-scale is a rapidly developing topic, in particular in superconductor-based hybrid devices. In this review paper, we first discuss the fundamental principles of electronic cooling in mesoscopic superconducting hybrid structures, the related limitations and applications. We review recent work performed in Grenoble on the effects of Andreev reflection, photonic heat transport, phonon cooling, as well as on an innovative fabrication technique for powerful coolers. 
\end{abstract}

\section{Introduction}

Let us consider a NIS tunnel junction made of a Normal metal and a Superconductor, in contact through an Insulator barrier. The superconductor's electronic density of states features an energy gap of width $\Delta$. Tunneling through the junction is possible for electrons with an energy $E$ (compared to the Fermi level $E_F$) such that $|E|>\Delta$. Only elastic processes are considered here. The charge current through such a junction biased at a voltage $V$ is \cite{TinkhamBook}:
\begin{eqnarray}                                                                                                                                                                                                                                                                                                                                                                                                                                                                                                                                                                                                                                   I=\frac{1}{eR_N}\int_{-\infty}^\infty n_S(E)[f_N(E-eV)-f_S(E)]dE\\
=\frac{1}{eR_N}\int_{0}^\infty n_S(E)[f_N(E-eV)-f_N(E+eV)]dE,
\label{eq:IV_theo}                                                                                                                                                                                                                                                                                                                                                                                                                                                                                                                                                                                                                                       \end{eqnarray}
where $R_N$ is the normal state resistance, $f_N$ the electron energy distribution in the normal metal and $n_S(E)$ the normalized BCS density of states in the superconductor. No Dynes parameter smearing the superconductor's density of states is taken into account. Expression (1) above is the usual one, while expression (2) can be obtained by a symmetry argument. The latter shows that the current does not depend on the superconductor temperature, only on the value of the gap.

When using it as a thermometer \cite{NahumAPL93}, a NIS junction is biased at a small and constant current $I_{th}$. The current can be adjusted so that the related heat flow (see below) can be safely neglected. The voltage drop $V_{th}$ in the thermometer junction pair is then measured and compared to its value calibrated against the bath temperature $T_{bath}$. This provides a measure of the electronic temperature $T_e$ with a sub-mK resolution. Let us note that the calibration is realized at equilibrium, with the superconductor and the normal metal being at the bath temperature. In contrast, practical experiments are usually conducted in quasi-equilibrium conditions where, to a first approximation, only the normal metal temperature changes. If one considers temperatures above about half the superconductor critical temperature $T_c/2$, the related decrease of the superconductor gap $\Delta$ has to be taken into account. The calibration then requires the full calculation of the voltage $V_{th}(T_e)$ using Eq.~(\ref{eq:IV_theo}).

The same tunnel current can cool down or heat up the normal metal electron population \cite{NahumAPL94,LeivoAPL96}. The heat current through a NIS junction reads:
\begin{equation}
\dot{Q}_N^0=\frac{1}{e^2R_N}\int_{-\infty}^\infty (E-eV)n_S(E)[f_N(E-eV)-f_S(E)]dE,
\label{eq:QN_theo}
\end{equation}
where $f_S$ is the energy distribution function in the superconductor. Equation \ref{eq:QN_theo} differs from Eq. 1 by an additional energy term in the integral, which also changes the symmetry compared to the charge current case: $\dot{Q}_N^0(-V)=\dot{Q}_N^0(V)$ while $I(-V)=-I(V)$. Equation \ref{eq:QN_theo}  describes both the cooling and heating regimes. With a voltage bias $V$ smaller than the gap $\Delta/e$, the tunnel current is selectively made out of high-energy electrons (or holes) from the tails of the distribution function. This cools the electronic population of the normal metal, which means that ${Q}_N^0$ is positive. Every tunnel event extracts a heat of about $k_BT$ while a heat $\Delta$ is dissipated in the system, so that the cooling efficiency $\dot{Q}_N^0/IV$ is about $k_B T_e/\Delta$. Above the gap ($|V| > \Delta/e$), the heat flow ${Q}_N^0$ becomes negative (the metal is heated). At large bias $|V| \gg \Delta/e$, the usual Joule heating regime is reached, ${Q}_N^0$ is equal to $- IV/2$. In every case, the full Joule power $IV$ is deposited in the device, so that a power $\dot{Q}_S=IV+\dot{Q}_N^0$ is transferred to the superconductor. 

In the cases studied here, we assume a quasi-equilibrium situation: the normal metal electrons and phonons follow thermal energy distribution functions at respective temperatures $T_e$ and $T_{ph}$. For electrons, this is justified because the inelastic scattering time (of the order of the phase-coherence time) is much shorter than the mean residence time estimated to about 100 ns for an island of typical dimension 50 nm $\times$ 1 $\mu$m$^2$ \cite{JLTP-Rajauria}. The thermalization of an isolated normal metal electron population to the thermal bath occurs through electron-phonon coupling. We will make use of the standard expression:
\begin{equation}
P_{e-ph}(T_{e},T_{ph})=\Sigma U(T_{e}^{5} - T_{ph}^{5}),
\end{equation}
of the electron-phonon coupling power, where $\Sigma$ is a material-dependent constant and $U$ is the metal volume. Several experiments have demonstrated the validity of this expression in mesoscopic (micron-sized) devices \cite{PRB-Urbina,JLTP-Godfrin}, although it is derived for a bulk material. In what follows, we take for Cu the established value $\Sigma$ = 2 nW.$\mu$m$^{-3}$.K$^{-5}$.

Although electronic cooling can be obtained in a single NIS junction where the normal metal is isolated from its electrical contact pad by the way of a thermally-insulating NS contact \cite{NahumAPL94}, the symmetric SINIS configuration with two NIS junctions acting as coolers brings the advantage of simplicity and a doubled cooling power \cite{LeivoAPL96}. Thanks to the current singularity at the gap edge, the SINIS geometry is rather immune to a small junctions' resistance asymmetry \cite{PhysicaPekola}. 

\begin{figure}[t]
\begin{center}
\includegraphics[width=0.9\linewidth]{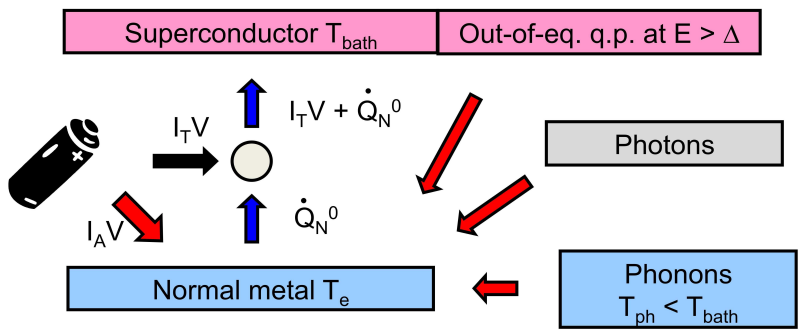}
\caption{Thermal diagram for electronic cooling: a heat $\dot{Q}_N^0$ is extracted from a normal metal N by running a current through a SINIS device, generating also a voltage V. The current is made of a single-particule tunnel current $I_T$ and an Andreev current $I_A$. Also shown are the different contributions discussed in this work that affect the actual cooling.}
\label{Scheme}
\end{center}
\end{figure}

The above considerations form the basis for electronic cooling. This topic has been discussed in recent reviews \cite{GiazottoRMP06,MuhonenRPP12}. Starting from a bath temperature of about 300 mK, an electronic temperature reduction by a factor of about 3 is routinely achieved in aluminum-copper hybrid devices that are biased at a voltage just below the gap of superconducting Al. The cooling power is in the 10 pW range for a tunnel junction area of about a square micron. Micro-coolers based on large lithographic junctions patterned on a membrane have demonstrated the cooling a bulk material \cite{ClarkAPL04} or a detector \cite{MillerAPL08}. Exciting new developments towards larger cooling powers \cite{NguyenNJP13} and cooling of macroscopic objects \cite{LowellAPL13} are emblematic of the present activity in the field.

In practice, electronic cooling in a normal metal is limited by several mechanisms of fundamental interest, see Fig. \ref{Scheme}. Electron-phonon scattering provides the main channel for thermalization of the electronic bath with respect to the substrate considered to be a thermal bath. The question of phonon cooling is therefore of primary importance. In recent years, we have also studied how other channels of thermalization contribute. In the following, we will discuss how Andreev current or photonic heat channel alter the electronic cooling in a SINIS device. We will not discuss quasi-particle relaxation in the superconducting electrodes, although it is a major and long-standing issue \cite{PekolaAPL00,RajauriaPRB09,PRB12-Rajauria}.

\section{Methods}

We study SINIS samples based on a normal metal island symmetrically coupled to two superconducting leads through tunnel barriers, see Fig. 2a. In general, they are prepared by electronic beam lithography on a resist bilayer, followed by two-angle evaporation in a high vacuum chamber \cite{CourtoisPRB95}. We focus on aluminium as a superconductor because of its unique high-quality oxide. As a thin film, its critical temperature $T_c$ is about 1.3 K. The normal metal is usually copper, with a diffusion constant $D$ of about 100 $cm^2/s$. The total normal-state resistance $R_n$ is in the range 2-3 k$\Omega$ for a tunnel junction area of about 1 $\mu m^2$. In addition to the two cooling junctions, one can add superconducting tunnel probes in order to probe the normal metal electronic temperature. In one case, we have built two such complete devices one on top of the other, in order to study the phonon bath behavior.

Low temperature measurements were performed in a $^3$He cryostat or a dilution cryostat. Filtering was provided by $\pi$-filters at room temperature and lossy micro-coaxial lines thermalized at the cold plate. Four-wire \mbox{d.c.} transport measurements were performed using home-made electronics combining independent current bias sources, one of them providing the voltage reference, the other being floating. The differential conductance $dI/dV(V)$ is obtained by numerical differentiation. We have taken special care to determine most accurately the sub-gap conductance of our current-biased samples down to a level of about $10^{-4}$ of the normal state conductance. Plots are usually a combination of multiple curves covering different measurement ranges, extending over several decades of current values. Semi-log plots are useful to provide a good understanding of the charge and heat transport. In such a plot, the sub-gap differential conductance of a NIS junction at constant temperature follows a linear behavior with the voltage drop. This comes from the fact that the Fermi distribution function of the energy can be approximated to a decaying exponential at energies well above or below the Fermi level.

In order to understand some of the experimental observations, we have investigated theoretically various aspects of heat and charge transport in NIS tunnel junctions. Since the devices measured experimentally are in the diffusive limit, the appropriate theoretical framework to describe thermo-electric transport is based on quasi-classical Keldysh-Usadel equations~\cite{Larkin86,Belzig99}. Specifically, using this formalism, we have calculated the two-particle current due to Andreev reflection~\cite{VasenkoPRB10}. This is relevant for the experiments discussed in Section~\ref{Andreev-heating} below. We also considered the effect of inelastic relaxation in the superconductor lead~\cite{VasenkoJLTP}. In the absence of strong relaxation, both the electric current and the cooling power at low voltages are suppressed. We could attribute this suppression to the effect of back-tunneling of non-equilibrium quasiparticles from the superconductor into the normal metal, a topic relevant for the experiments described in Section~\ref{phonon-cooling} below.

\section{Heating induced by the Andreev current}
\label{Andreev-heating}

Operation and limitations of NIS electronic coolers in the low-temperature regime ($T \ll T_c$) have been questioned in the past \cite{PRL-Pekola}. A residual smearing of the superconductor density of states has been proposed as the origin of the inferior performance with respect to theoretical expectations.

At low energy $|E|<\Delta$, charge transfer at a NIS interface occurs through Andreev reflection \cite{Andreev,SaintJames}. In the normal metal, an electron (a hole) impinging on the superconducting interface is reflected as a hole (an electron), enabling the transfer of a Cooper pair into (out of) the superconductor. For a NIS tunnel junction of intermediate or low transparency, the Andreev reflection probability is predicted to be vanishingly small. Taking into account the quasi-particles confinement in the vicinity of the interface, this is no longer true. This confinement can be induced by  disorder or the presence of a second barrier in the normal metal. A single quasiparticle then experiences several collisions with the interface \cite{PRL-VanWees,PRL-Hekking} so that the actual Andreev reflection transmission coefficient corresponds to the coherent addition of many individual transmission probabilities. Therefore, the Andreev sub-gap current can significantly exceed the ballistic case prediction \cite{PRL-Kastalsky}.

\begin{figure}[t]
\begin{center}
\includegraphics[width=\linewidth]{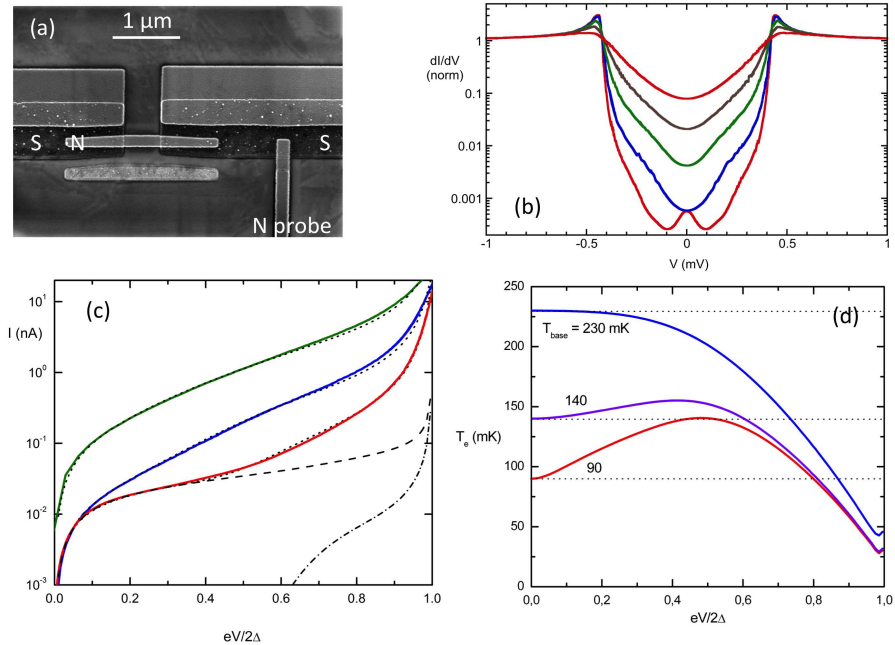}
\caption{(a) Scanning electron microscope (SEM) micrograph of a typical cooler sample made of a normal metal Cu electrode (light grey) connected to two superconducting Al reservoirs (dark grey) through tunnel junctions. One additional probe junction connected to one Al reservoir is visible at the bottom.
(b) Normalized differential conductance as a function of the voltage and at the cryostat temperatures of 90 (red curve), 230 (blue), 330 (green), 440 mK (brown) and 600 mK (red).
(c) Current-voltage characteristic of the S-I-N-I-S cooler junction as a function the voltage at cryostat temperature of 90 (red line), 230 (blue) and 330 mK (green) together with best-fit calculated curves. Dot-dashed line: model including the cooling by the tunnel current, with parameters 2$\Delta$ = 0.43 meV, K.A = 144 $W.K^{-4}$. Dashed line: model including in addition the Andreev current, but not the related heating, with the parameters $D=$ 80 $cm^2.s^{-1}$, $L_\varphi=$1.5 $\mu$m. Dotted lines: full model taking into account the Andreev current and the related heating. Compared to theoretical calculation, the Andreev current was multiplied by 1.55.
(d) Dependence of the calculated electronic temperature with the voltage as derived from the fit of the experimental results and for a series of cryostat temperatures: 90 (red symbols), 140 (green) and 230 mK (blue).}
\label{Andreev}
\end{center}
\end{figure}

Fig. \ref{Andreev}b displays on a logarithmic scale the differential conductance of a SINIS device for a series of cryostat temperatures \cite{RajauriaPRL08}. We observe in the high temperature ($T >$ 200 mK) data an upward curvature which constitutes a clear signature of electronic cooling. In the low temperature regime ($T <$ 200 mK), clearly a different characteristic is obtained with a differential conductance peak at zero bias. The zero-bias differential conductance increases while the temperature is lowered below about 200 mK. We ascribe the low bias differential conductance peak to an Andreev current, i.e. a double particle tunnel current created by Andreev reflections at the NIS junctions.

In order to calculate the Andreev current, we take into account the finite gap $\Delta$ and the disorder both in the normal metal and in the superconductor. We consider the 1D regime where the coherence length of an Andreev pair in the normal metal $L_{E}=\sqrt{\hbar D/E}$ \cite{Superlattices-Courtois} is much larger than the junction dimension. Usadel equations formalism can also be used and provide the same result \cite{VasenkoPRB10}. In the probe junction, the electronic temperature can be considered as constant and very close to the cryostat temperature. In this case, the isothermal calculation fits nicely the experiment (not shown). In the cooler junction, the non-linear behavior in a semi-log scale (Fig. \ref{Andreev}c) shows that electronic cooling (and heating) has to be taken into account.

The work performed by the current source feeding the circuit with a (Andreev) current $I_{A}$ generates a Joule heating $I_{A}V(>0)$ that is deposited in the normal metal. The full heat balance equation for the normal metal electrons can then be written as:
\begin{equation}
2\dot{Q}_N^0+P_{e-ph}-I_{A}V=0.
\end{equation}
Here the factor 2 accounts for the presence of two tunnel junctions. In the low-temperature limit considered here, the normal metal phonons can be considered as thermalized at the bath temperature: $T_{ph}=T_{bath}$. With this complete heat balance equation taken into account, we calculate the current-voltage characteristic at every cryostat temperature. The agreement is very good at every temperature, see Fig. \ref{Andreev}c, and covers 4 orders of magnitude for the current. At temperatures above about 300 mK, phonon cooling introduces significant corrections, see below.

The numerical solution of the heat balance equations also provides the electron temperature at every bias. Fig. \ref{Andreev}d shows the calculated dependence of the normal metal central island electron temperature with the voltage across the cooler. At very low temperature, the electron temperature first increases with the bias due to Andreev current heating, before decreasing due to the single quasi-particle tunnel current-based cooling effect. Although the Andreev current is a small effect in a such junction if one considers the charge current, this is no longer true if one considers the heat current. Compared to a Joule power $IV$, the Andreev current contributes fully to heating while the tunnel current cools with a moderate efficiency $k_BT_e/\Delta$ of about 5 $\%$ at a 100 mK electron temperature.

\section{Photonic heat channel}
\label{photon}

In metallic systems, heat conduction can be achieved by electrons, phonons and also photons. The photonic channel~\cite{Schmidt04} was recently revealed experimentally \cite{Nature-Meschke,PRL-Timofeev-1} at very low temperature in devices including superconducting transmission lines. With a good matching between the source and the drain, the thermal conductance of a superconducting transmission line is equal to the thermal conductance quantum \cite{Pendry83}: $K_Q =  k_{B}^{2}T\pi / 6 \hbar$. The photonic channel for heat transfer can in principle couple metallic systems that are galvanically isolated, \mbox{e.g.} through a capacitor, or weakly coupled, \mbox{e.g.} through a tunnel junction. This effect can be beneficial in some cases, but also detrimental when one wants to maintain two electronic populations at different quasi-equilibrium temperatures. This is precisely the case encountered in electronic coolers.

We have investigated theoretically the photonic heat transfer through a general reactive impedance, \mbox{i.e.} a linear coupling circuit that contains a capacitor, an inductance, a resonant circuit or a transmission line \cite{PascalPRB11}. Only the case of a capacitor will be discussed here, but the same approach can be used in all cases. We follow a simple circuit approach~\cite{Beenakker03}, valid at low temperatures when the relevant photons have wavelengths larger than the size of the typical circuit element. The metallic parts can then be treated as lumped elements characterized by an electrical impedance.

Let us consider a circuit of two pure resistors 1 and 2 of resistance $R$ coupled capacitively through an impedance $Z_c=1/i \omega C$, see Fig. 3b inset. With a small temperature difference and using the dimensionless frequency
\begin{equation}
x=\frac{\hbar \omega}{2k_BT},
\end{equation}
the exchanged power $P_{1,2}$ and the thermal conductance $K_C$ are given by:
\begin{equation}
P_{1,2} = K_C \Delta T = \frac{T k_{B}^{2}}{\pi\hbar} \int_{0}^{\infty}\emph{d}\textrm{x}{\cal T_C}(\textrm{x}) \frac{\textrm{x}^{2}}{\sinh^{2}(\textrm{x})} \Delta T.
\label{conductance}
\end{equation}
taking into account the frequency-dependent photon transmission coefficient:
\begin{equation}
{\cal T}_C(\textrm{x}) = \frac{\textrm{x}^{2}}{\textrm{x}^{2}+\alpha_C^{2}}.
\label{filtertrans}
\end{equation}
The cross-over frequency is determined by the parameter
\begin{equation}
\alpha_C=\frac{\hbar}{4 R C k_{B} T};
\end{equation}
it separates a low-frequency regime $\textrm{x} \ll \alpha_C$ where the capacitor is opaque and ${\cal T}_C(\textrm{x}) \sim \textrm{x}^2/\alpha_C^{2} \ll 1$ from a high-frequency regime $\textrm{x} \gg \alpha_C$ where the capacitor is transparent and ${\cal T}_C(\textrm{x}) \sim 1$.

\begin{figure}[t]
\centering
\includegraphics [width=1\linewidth]{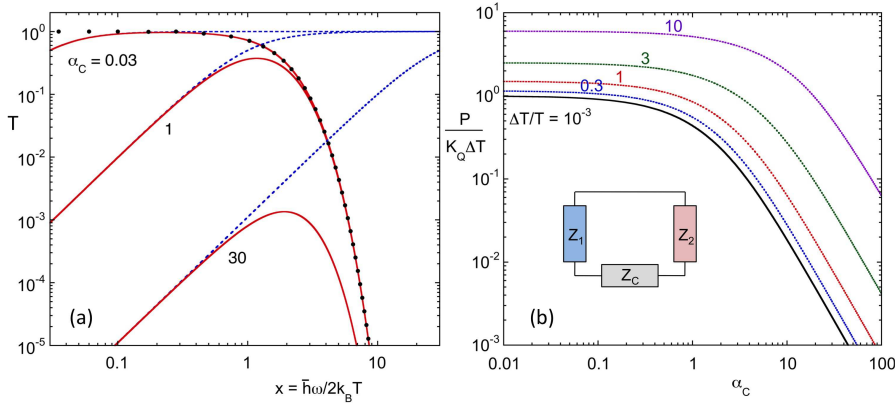}
\caption{(a) Case of a capacitive coupling. Normalized spectrum of the thermal noise power density $x^2/\sinh^2 x$ (black dots), the photon transmission coefficient ${\cal T_C}(x) = x^2/(x^2+\alpha_C^2)$ (dotted blue line) and of photonic heat (thin red full line) as a function of the dimensionless frequency, for values of the parameter $\alpha_C$ = 0.03, 1 and 30 from top to bottom. The frequency is plotted in units of the thermal frequency $2k_BT/\hbar$. We consider the case of perfect resistance matching $R_1 = R_2 =R$. (b) Dependence of the photonic power through a capacitive coupling impedance on the parameter $\alpha_C$ for different values of the relative temperature difference $\Delta T/T$, in units of $K_Q \Delta T$, the maximum photonic power in the case of a linear response.}
\label{Photon}
\end{figure}

Fig.~\ref{Photon}a displays the photon transmission coefficient ${\cal T}_C$. The limit $\alpha_C \ll 1$ means that the capacitance is large, i.e. it has a negligible impedance over most of the thermal spectrum. The transparency ${\cal T}_C$ is then equal to unity and one recovers $K_C = K_Q$. In the limit $\alpha_C \gg 1$, the photonic signal is strongly suppressed by the RC filter composed of the series capacitance and the receiver resistance, leading to $K_C \ll K_Q$.

The total photonic power, integrated over the full frequency range, is shown in Fig. \ref{Photon}b as a function of the parameter $\alpha_C$. We compare both linear and non-linear response, changing the values of the relative temperature difference $\Delta T/T$. The total power is maximal for small $\alpha_{C}$; it decays as  $1/\alpha_{C}^2$ when $\alpha_{C}$ is large. A cross-over between the linear regime $P \propto \Delta T$ and the non-linear regime occurs at $\Delta T/T \approx 1$. When the temperature difference is large, the photonic thermal conductance is larger than its quantum because of the broader frequency range of the emitted photons. Nevertheless, the thermal conductance remain of the order of $K_Q$, \mbox{i.e.} 1 pW/K at 1 K, and thus little contributes in usual situations.

\section{Phonon cooling}
\label{phonon-cooling}

As mentioned above, the electron-phonon coupling provides the main channel for coupling of the cooled electron population to the thermal bath of the cryostat. The question arises of the actual temperature of the phonon bath in the normal metal. At a temperature $T$, phonons have a thermal wavelength of the order of $hc/k_BT$, where $c$ is the material-dependent sound velocity. This dominant wavelength amounts to about 200 nm in Cu at 1 K, which is the order of magnitude of an usual device dimensions. It is thus generally assumed that phonons in a mesoscopic quantum device are strongly mixed with the substrate phonons and are thus thermalized at the bath temperature \cite{GiazottoRMP06}.

Nevertheless, phonon cooling is at the heart of the possibility of cooling a bulk detector \cite{ClarkAPL05} or a quantum device \cite{VercruyssenAPL11} supported on a membrane cooled by superconducting tunnel junctions. The same situation also holds for suspended metallic beams \cite{KoppinenPRL09,MuhonenAPL09}. Recently, measurements of the electron-phonon coupling strength in a thin metallic film at T $\approx$ 0.1-0.3 K demonstrated that it is nearly completely substrate-insensitive \cite{UnderwoodPRL11}. This supports the idea of an independent phonon population in the metallic thin film. While this phonon bath could exhibit specific properties due to its reduced dimensionality \cite{QuPRB05,HekkingPRB08}, only small deviations from bulk material properties were observed in suspended devices \cite{KarvonenPRL07}.

\begin{figure}[t]
\begin{center}
\includegraphics[width= \linewidth]{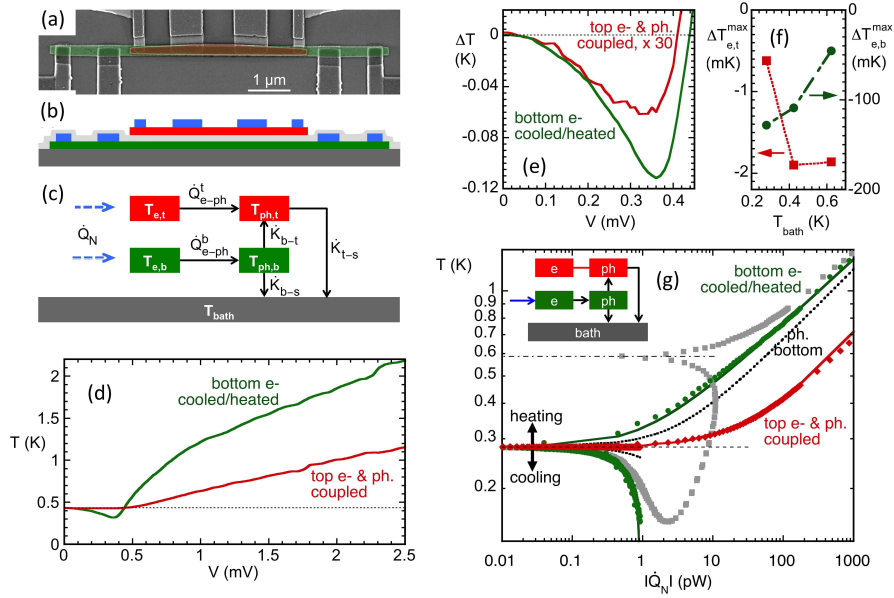}
\caption{(a) SEM image of the device. Each of the two normal islands (colorized in red or in green) is inserted between two sets of superconducting junctions. One junction pair is used as a cooler (or heater) and the other one is used as thermometer.
(b) Schematic side-view of the set-up. The top (t) and bottom (b) levels are galvanically isolated from each other by a 40 nm thick layer of Si.
(c) Heat transfer model. Electrons and phonons of the bottom (top) island at a respective temperature $T_{e,b}$ and $T_{ph,b}$ ($T_{e,t}$ and $T_{ph,t}$) exchange a heat power $\dot{Q}_{e-ph}^{b}$ ($\dot{Q}_{e-ph}^{t}$). Phonons of each island are coupled together via the Kapitza power $\dot{K}_{b-t}$, and to the bath phonons via $\dot{K}_{b-s}$ and $\dot{K}_{t-s}$.
(d) Electronic temperatures $T_{e,b}$ and $T_{e,t}=T_{ph,t}$ as a function of the voltage $V$ across the bottom cooler/heater junctions at a bath temperature $T_{bath}=432$ mK.
(e) Temperature variations $\Delta T_{e,b}$ and $\Delta T_{e,t}=\Delta T_{ph,t}$ based on the same data, the latter temperature change being amplified by a factor 30.
(f) Maximum of $\Delta T_{e,b}$ (circles) and $\Delta T_{e,t}$ (squares) as a function of the bath temperature.
(g) Electronic temperatures $T_{e,b}$ (green circles) and $T_{e,t}$ (red diamonds) as a function of the absolute value of the power injected in the bottom island. Grey squares show $T_{e,b}$ plotted as a function of the raw power $\dot{Q}_N^0$, while all other data are plotted as a function of the corrected power $\dot{Q}_N=\dot{Q}_N^0+\alpha Q_S$. The dash-dotted and the dotted lines corresponds to the point where the power $\dot{Q}_N^0$ or $\dot{Q}_N$ respectively change sign. The full lines are fits calculated using a single set of parameters. The black dotted lines indicate the calculated phonon temperature of the island that is cooled or heated. Bath temperature is 281 mK.}
\label{Phonon}
\end{center}
\end{figure}

We have devised a dual-level sample with two independent but stacked SINIS devices including electron thermometers, see Fig. \ref{Phonon}a. The experiment \cite{PascalPRB13} consists in current-biasing one of the two level's cooler junction pair while monitoring simultaneously the related voltage drop $V$ as well as the two levels' electronic temperatures $T_{e,b}$ and $T_{e,t}$, where "b" stand for bottom and "t" for top. As no power is directly injected in the unbiased electronic bath, its temperature is equal to the included metal's phonon temperature. For voltages below $2\Delta/e$ applied to the bottom level, we observe the expected electronic cooling, see Fig. \ref{Phonon}d. At voltages $V$ above $2\Delta/e$, we observe a hot-electron regime: the temperature $T_{e,b}$ increases and goes well above the bath temperature of 432 mK.

Remarkably, when the bottom level electronic temperature decreases, the top electronic temperature $T_{e,t}$ also diminishes with a variation $\Delta T_{e,t}$ reaching a maximum of - 2.0 mK, see Fig.~\ref{Phonon}e. As the operation of the electronic cooler is dissipative as a whole, \mbox{i.e.} heat is dissipated in the chip, this observation cannot be related to an improper thermalization of the chip or of electrical leads. The observed cooling of the top level therefore demonstrates phonon cooling in the normal conductors of the experiment.

The cooling/heating power was calculated by using Eq.~\ref{eq:QN_theo}, the measured electronic temperature, and a value $\Delta=$ 214 $\mu eV$. As a fraction of the hot quasi-particles injected in the superconductor tunnels back in the normal metal, we have described this as a correction to the power proportional to $\dot{Q}_S$ \cite{UllomPhysica00}: the net power then writes $\dot{Q}_N=\dot{Q}_N^0 +\alpha \dot{Q}_S$. The fit parameter value $\alpha$ = 0.087 is comparable to what appears in the literature \cite{UllomPhysica00}. The electronic temperature as a function of the net power $\dot{Q}_N$ absolute value then follows a single curve when one goes through the maximum cooling point, see Fig.~\ref{Phonon}g.

We consider a thermal model, see Fig.~\ref{Phonon}c, assuming two distinct phonon populations at quasi-equilibrium at temperatures $T_{ph,b}$ and $T_{ph,t}$ in the bottom and top metallic islands respectively. We assume that two neighbouring phonon populations (here x and y) are coupled through a Kapitza heat flow of the form
\begin{equation}
\dot{K}_{x-y}=k_{xy}a_{xy}[T_{ph,x}^4-T_{ph,y}^4]
\end{equation}
\cite{GiazottoRMP06}, where $a_{xy}$ holds for the contact area between the two considered populations and $k_{xy}$ is an interface materials-dependent parameter. We have chosen to take a common value for the substrate-bottom and bottom-top Kapitza parameters. The fit-derived value $k_{bt}=k_{bs}$ = 45 pW.$\mu$m$^{-2}$.K$^{-4}$ compares well to values from the literature for a Cu-Si interface \cite{SwartzRMP89}. As for the top-substrate coupling, considering the contact area $a_{ts}$ to be the area of the tunnel junctions connected to the island, one obtains a Kapitza coefficient of 920 pW.$\mu$m$^{-2}$.K$^{-4}$, which is much larger than anticipated. Thus the heat transfer occurs presumably also along the continuous Si layer separating the two levels, which explains the relatively modest amplitude of the observed phonon cooling.

From the thermal balance relations, one can calculate the phonon temperature variation in the cooled or heated metal, see dotted lines in Fig.~\ref{Phonon}g. The phonon population temperature decoupling is significant in the temperature range 0.3 - 1 K, consistently with previous estimates \cite{RajauriaPRL07}. At lower temperature, the electron-phonon bottle-neck makes the phonon temperature tend to the bath temperature. At a bath temperature below 100 mK, phonon cooling becomes negligable.

Our experimental study thus demonstrates the existence of an independent phonon bath in a quantum device. The thermal couplings are well described with the usual laws for electron-phonon coupling and Kapitza resistances. This new understanding has significant outcomes in the analysis of quantum nano-electronic devices thermal behavior.

\section{Towards large electronic cooling power}

Optimizing a SINIS device for electron cooling can be achieved by increasing the heat current and/or isolating better the cooled electron bath. As the heat current is proportional to the tunnel barriers' conductance, reducing the barriers' thickness is the first option. Nevertheless, this can lead to the appearance of two-particle Andreev reflection processes at low energy, which deposit heat in the normal metal \cite{RajauriaPRL08}. The obvious alternative is to increase the junction area at a fixed transparency \cite{ONeilPRB12}. We present a method to fabricate large area SINIS devices of high quality and with a suspended normal metal \cite{NguyenAPL12}. The method is based on a pre-deposited multilayer of metals, which can be prepared at the highest quality. The upper normal part is suspended in the first lithography, which keeps it isolated from the substrate. The second lithography defines the junction area with any geometry of interest.

\begin{figure}[t]
\begin{center}
\includegraphics[width=\columnwidth,keepaspectratio]{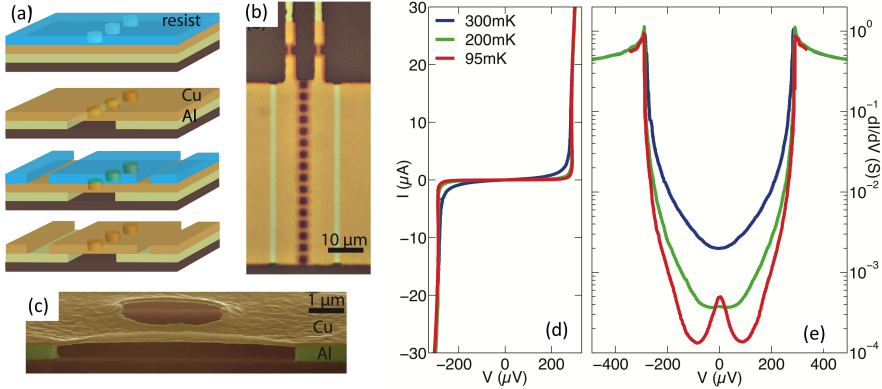}
\caption{(a) From top: fabrication starts with an Al/AlO$_x$/Cu multilayer, on which a photoresist is patterned with contact pads and holes. Then, Cu and Al are successively etched, leaving a suspended membrane of Cu along the line of adjacent holes. A second lithography and etch define the Cu central island. (b) Optical microscope image showing regions by decreasing brightness: bare Al, Cu on Al, suspended Cu and substrate. On the top, two thermometer junctions are added.
(c) Colorized scanning electron micrograph of a sample cut using Focused Ion Beam, showing the Cu layer suspended over the holes region. The thickness of Al and Cu is 400 and 100 nm, respectively.
(d) Current-voltage characteristic and (e) differential conductance of a sample at different cryostat temperatures.}
\label{Large}
\end{center}
\end{figure}

The fabrication starts with depositing a Al/AlO$_x$/Cu multilayer on an oxidized silicon substrate. A first deep ultra-violet lithography is used to define the overall device geometry. The central part is a series of adjacent holes of diameter 2 $\mu$m and with a side-to-side separation of 2 $\mu$m, see Fig. \ref{Large}b. The copper layer is etched away over the open areas using either Ion Beam Etching (IBE) or wet etching. The aluminum is then etched through the same resist mask, using a weak base. The etching time is controlled as to completely remove aluminum from the circular region within a horizontal distance of about 2 $\mu$m starting from the hole side. The line of adjacent holes visible in Fig. \ref{Large}b therefore creates a continuous gap in the Al film, bridged only by a stripe of freely hanging Cu. The area of the NIS junctions is defined by a second lithography. Through the open areas, trenches are etched into the copper layer only, using one of the two methods cited above. These trenches allow to isolate a copper island, which forms the central normal metal part of the SINIS device.

Fig. \ref{Large}d,e shows the current-voltage characteristic and the numerically-derived differential conductance of a typical sample at various cryostat temperatures. We have measured the electronic temperature as a function of the cooler voltage bias by using the two attached smaller junctions, see Fig. \ref{Large}b, as an electron thermometer. At a bath temperature $T_{bath}$ of 300 mK and at the optimum bias point, the measured electronic temperature $T_e$ reaches a minimum of 240 mK.

Removing the contact between the substrate and the cooled metal by suspending the latter is quite promising for electronic refrigeration applications as it can significantly improve cooling of electrons and phonons. Our approach also has the advantage that as fabrication starts with preparing the multilayer, the wafer can be baked in ultra-high vacuum environment, which is an essential ingredient for obtaining pinhole-free large-area NIS junctions. This process has recently been improved by including a quasi-particle drain in good contact with the superconducting electrodes. A nanoWatt cooling power  \cite{NguyenNJP13} as well as a base temperature down to 30 mK have been demonstrated \cite{NguyenUnpublished}. This high performance can be understood only by assuming some phonon cooling in the suspended metal.

\section{Conclusion}

As a summary, we have reviewed a series of experimental studies combined with theoretical analyses of different phenomena at work in SINIS electronic coolers. Andreev current heating, coupling to phonons, phonon cooling, photonic coupling are all important limiting factors and either can dominate, depending on the practical case. The field is in constant progress, for example with the demonstration of improved performance \cite{NguyenNJP13} and suitability for the cooling of macroscopic objects \cite{LowellAPL13}.

\begin{acknowledgements}
This work was funded by the EU FRP7 low temperature infrastructure MICROKELVIN and by Institut universitaire de France. Samples were fabricated at Nanofab platform - CNRS. This work has been led in collaboration with S. Rajauria, L. M. A. Pascal, A. Fay, B. Pannetier, A. Vasenko, T. Crozes and T. Fournier. We acknowledge fruitful discussions along the years with J. P. Pekola, F. Giazotto and M. Meschke.
\end{acknowledgements}


\end{document}